\documentstyle[12pt]{article}

\input epsf

\ifx\epsffile\undefined
\newlength{\epsfysize}
\def\epsffile#1{}
\else
\fi

\catcode`\@=11
\long\def\@makefntext#1{
\protect\noindent \hbox to 3.2pt {\hskip-.9pt
$^{{\ninerm\@thefnmark}}$\hfil}#1\hfill}		

 \def\@makefnmark{\hbox to 0pt{$^{\@thefnmark}$\hss}}  

\def\ps@myheadings{\let\@mkboth\@gobbletwo
\def\@oddhead{\hbox{}
\rightmark\hfil\ninerm\thepage}
\def\@oddfoot{}\def\@evenhead{\ninerm\thepage\hfil
\leftmark\hbox{}}\def\@evenfoot{}
\def\sectionmark##1{}\def\subsectionmark##1{}}

\newcounter{sectionc}\newcounter{subsectionc}\newcounter{subsubsectionc}
\renewcommand{\section}[1] {\vspace{0.6cm}\addtocounter{sectionc}{1}
\setcounter{subsectionc}{0}\setcounter{subsubsectionc}{0}\noindent
	{\bf\thesectionc. #1}\par\vspace{0.4cm}}
\renewcommand{\subsection}[1] {\vspace{0.6cm}\addtocounter{subsectionc}{1}
	\setcounter{subsubsectionc}{0}\noindent
	{\it\thesectionc.\thesubsectionc. #1}\par\vspace{0.4cm}}
\renewcommand{\subsubsection}[1]
{\vspace{0.6cm}\addtocounter{subsubsectionc}{1}
	\noindent {\rm\thesectionc.\thesubsectionc.\thesubsubsectionc.
	#1}\par\vspace{0.4cm}}

\newcounter{appendixc}
\newcounter{subappendixc}[appendixc]
\newcounter{subsubappendixc}[subappendixc]

\renewcommand{\appendix}[1] {\vspace{0.6cm}
        \refstepcounter{appendixc}
        \setcounter{figure}{0}
        \setcounter{table}{0}
        \setcounter{equation}{0}
        \renewcommand{\thefigure}{\Alph{appendixc}.\arabic{figure}}
        \renewcommand{\thetable}{\Alph{appendixc}.\arabic{table}}
        \renewcommand{\theappendixc}{\Alph{appendixc}}
        \renewcommand{\theequation}{\Alph{appendixc}.\arabic{equation}}
        \noindent{\bf Appendix \theappendixc #1}\par\vspace{0.4cm}}

\def\abstracts#1{{
	\centering{\begin{minipage}{30pc}\tenrm\baselineskip=12pt\noindent
	\centerline{\tenrm ABSTRACT}\vspace{0.3cm}
	\parindent=0pt #1
	\end{minipage}}\par}}

\renewenvironment{thebibliography}[1]
	{\begin{list}{\arabic{enumi}.}
	{\usecounter{enumi}\setlength{\parsep}{0pt}
\setlength{\leftmargin 1.25cm}{\rightmargin 0pt}
	 \setlength{\itemsep}{0pt} \settowidth
	{\labelwidth}{#1.}\sloppy}}{\end{list}}

\newcounter{itemlistc}
\newcounter{romanlistc}
\newcounter{alphlistc}
\newcounter{arabiclistc}

\newcommand{\fcaption}[1]{
        \refstepcounter{figure}
        \setbox\@tempboxa = \hbox{\tenrm Fig.~\thefigure. #1}
        \ifdim \wd\@tempboxa > 5.5in
           {\begin{center}
        \parbox{5.5in}{\footnotesize\baselineskip=12pt Fig.~\thefigure. #1}
            \end{center}}
        \else
             {\begin{center}
             {\tenrm Fig.~\thefigure. #1}
              \end{center}}
        \fi}

\newcommand{\tcaption}[1]{
        \refstepcounter{table}
        \setbox\@tempboxa = \hbox{\tenrm Table~\thetable. #1}
        \ifdim \wd\@tempboxa > 6in
           {\begin{center}
        \parbox{6in}{\tenrm\baselineskip=12pt Table~\thetable. #1}
            \end{center}}
        \else
             {\begin{center}
             {\tenrm Table~\thetable. #1}
              \end{center}}
        \fi}

\def\fnt#1#2{\footnotetext{\kern-.3em
	{$^{\mbox{\sevenrm #1}}$}{#2}}}

 1
 1
 1

\font\tenbf=cmbx10
\font\tenrm=cmr10
\font\tenit=cmti10

\font\ninerm=cmr9

\begin{document}

\centerline{\tenbf DYNAMICAL SUPERSYMMETRY BREAKING}
\baselineskip=16pt
\centerline{\tenbf IN SUPERGRAVITY THEORIES}
\vspace{0.8cm}
\centerline{\tenrm JONATHAN~BAGGER}
\baselineskip=13pt
\centerline{\tenit Department of Physics and Astronomy}
\baselineskip=12pt
\centerline{\tenit The Johns Hopkins University}
\centerline{\tenit Baltimore, MD  \ 21218,\ \ USA}
\vspace{0.9cm}
\abstracts{
In rigid supersymmetry,
generic models of dynamical supersymmetry breaking contain a light Goldstone
boson, called the $R$ axion.  We show that supergravity effects explicitly
break the $R$ symmetry and give mass to the $R$ axion.  For visible and
renormalizable hidden sector models, the massive $R$ axion is free from
astrophysical and cosmological problems.  For nonrenormalizable hidden
sector models, the $R$ axion suffers from cosmological difficulties
similar to those of the moduli fields in string theory.}
\vfil

\def\sm{SU(3) $\times$ SU(2) $\times$ U(1)~}
\def\E{{\cal E}}
\def\D{{\cal D}}
\def\V{{\cal V}}
\def\bD{\bar{\cal D}}
\def\btheta{\bar\theta}
\def\roughly#1{\raise.3ex\hbox{$#1$\kern-.75em\lower1ex\hbox{$\sim$}}}
\newcommand{\nc}{\newcommand}
\nc{\lsim}{\begin{array}{c}\,\sim\vspace{-21pt}\\< \end{array}}
\nc{\gsim}{\begin{array}{c}\sim\vspace{-21pt}\\> \end{array}}

\rm\baselineskip=14pt

\section{Introduction}
\vspace{-0.25cm}

Supersymmetry has long been viewed as an attractive candidate for
physics beyond the standard model.  There are many reasons for this,
including the fact that supersymmetry stabilizes the gauge hierarchy
against radiative corrections.  In a supersymmetric theory, once the
weak scale is fixed to be much smaller than the Planck scale, the
hierarchy $M_W \ll M_P$ is preserved by radiative corrections.

In fact, the hierarchy can be preserved even if supersymmetry is broken.
This follows from the fact that divergent radiative
corrections are cut off by the scale of supersymmetry breaking.
The hierarchy is preserved if the radiative corrections obey the
following naturalness condition,
\begin{equation}
\delta M^2_W\ \simeq\ \eta^2 M^2_S\ \simeq\ M^2_W\ ,
\end{equation}
where $M_S$ is the scale of supersymmetry breaking and $\eta$ is an
effective coupling that parametrizes the strength with which the
supersymmetry breaking is communicated to the everyday world.

Although supersymmetry stabilizes the hierarchy, it does not necessarily
explain the origin of the ratio
\begin{equation}
{M_W \over M_P}\ \simeq\ 10^{-17}\ .
\end{equation}
This motivates one to consider models of dynamical symmetry breaking,
where supersymmetry (and electroweak symmetry) are dynamically broken.
In such models all scales arise from $M_P$ through
dimensional transmutation,
\begin{equation}
\Lambda\ \simeq\ \exp\left({-8\pi^2\over g^2}\right)\,M_P\ .
\end{equation}
Supersymmetry is unbroken to all orders of perturbation theory, but
nonperturbative effects generate a ground state that breaks
supersymmetry \cite{dsb,ads}.

Models with dynamical supersymmetry breaking offer a natural explanation
for the origin of the gauge hierarchy.  Typically, such theories contain
two sectors.  One contains the minimal supersymmetric standard model (or
some simple extension).  The other dynamically breaks supersymmetry.
The full theory can then be classified by the way in which supersymmetry
breaking is transmitted to the fields of the standard model \cite{banks
kaplan nelson}:

\begin{itemize}
\item
{\it Visible Sector Models.}  In VS models, supersymmetry breaking is
communicated by gauge interactions.  Supersymmetry is broken by the
dynamical theory at the scale $\Lambda$, so $ M^2_S \simeq \Lambda^2$.
The effective coupling $\eta$ is of order $(g/4\pi)^n$, where $n$ counts
the number of loops necessary to connect the two sectors.  In such models,
$\Lambda$  is typically of order $10^5$ GeV.

\item
{\it Renormalizable Hidden Sector Models.}  In RHS models, supersymmetry
breaking is transmitted by gravitational interactions.  Supersymmetry is
still broken by a dynamical theory at the scale $\Lambda$,
so $M^2_S \simeq \Lambda^2$.  Now, however, $M_S^2 \simeq M_W M_P$, so
$\Lambda \simeq 10^{10}$ GeV.  In these models, the effective coupling
is much smaller, $\eta^2 \simeq M_W/M_P$.

\item
{\it Nonrenormalizable Hidden Sector Models.}  In NRHS models, the
supersymmetry breaking is communicated by gravity, so $M_S^2
\simeq M_W M_P$.  The difference is that supersymmetry is not broken
in the limit $M_P \rightarrow \infty$.  Typically,  $M^2_S \simeq
\Lambda^3/M_P$, which implies $\Lambda \simeq 10^{13}$ GeV.  NRHS
models are not renormalizable in the sense that supersymmetry breaking
relies on nonrenormalizable operators suppressed by powers of $1/M_P$.
As in the previous case, $\eta^2 \simeq M_W/M_P$.
\end{itemize}

In each of these cases, the dynamical supersymmetry breaking has important
consequences for the everyday world.  These effects can be summarized
in terms of a spurion in an effective superspace lagrangian.
Let us examine each type of model in turn.

\subsection{Visible Sector Models}
\vspace{-0.25cm}

In VS models, the supersymmetry breaking can be understood
in terms of a \sm chiral superfield $S$.  Below the scale $\Lambda$,
this superfield appears in the effective lagrangian through terms of
the form
$${\eta^2 \over \Lambda^2}\ \int d^4\theta\ S^+S\,\Phi^+\Phi$$
\begin{equation}
{\eta \over \Lambda}\ \int d^2\theta\ S\,W^\alpha W_\alpha\ ,
\end{equation}
where $\Phi$ and $W^\alpha$ are matter and gauge superfields of the
supersymmetric standard model.  The field $S$ is a spurion, so it
has no dynamics, just an expectation value $\langle S \rangle
\simeq M^2_S\,\theta\theta$.  Therefore these terms give rise to
soft supersymmetry-breaking masses of order $\eta M_S$ for the
squarks, sleptons and gauginos.

In VS models, the soft masses are, in principle, calculable.  However,
most models are quite complicated, with the soft masses arising at
multi-loop order.  (See, for example, \cite{dine nelson}.  For
recent progress, see \cite{dine nelson shirman}.)  Therefore the effective
coupling $\eta$ tends to be small, perhaps of order $10^{-3}$, in which
case $M_S \simeq \Lambda \simeq 10^5$ GeV.

\subsection{Renormalizable Hidden Sector Models}
\vspace{-0.25cm}

In RHS models, supersymmetry breaking is communicated by the supergravity
auxiliary fields.  The superdeterminants of the superspace vielbeins play the
role of the spurions,
$$\int d^4\theta\ E\,\Phi^+\Phi$$
\begin{equation}
\int d^2\theta\ \E\,W^\alpha W_\alpha\ ,
\label{sg-spurion}
\end{equation}
where we use the notation of \cite{wess bagger}.
The normalizations of these terms are fixed by the normalizations of the
superspace kinetic energies.

When supersymmetry is broken, the fields $E$ and $\E$ develop expectation
values,
$$\langle E \rangle \ \simeq \ 1\ -\ {M^2_S \over M_P}\,\theta\theta\ -
\ {M^2_S \over M_P}\,\btheta\btheta\ +\ {M^4_S \over M^2_P}\,
\theta\theta\btheta\btheta $$
\begin{equation}
\langle \E \rangle \ \simeq \ 1\ -\ 3\,{M^2_S \over M_P}\,\theta\theta\ .
\end{equation}
The spurion $E$ induces universal masses of order $M^2_S/M_P \simeq M_W$
for the squarks and the sleptons.  Therefore, in these models, $\Lambda
\simeq M_S \simeq 10^{10}$ GeV.

As in the VS case, RHS models have the advantage that they are,
in principle, calculable.  The problem is that (\ref{sg-spurion})
gives a vanishing tree-level gaugino mass. (For two viewpoints on
the phenomenology of light gluinos, see
\cite{massless gaugino}.)  This follows from the fact that the
supergravity auxiliary field that appears in $\E$ also appears in
the supersymmetric gauge field strength $W_\alpha = (\bD\bD - 8 R)
e^{-V} \D_\alpha e^{V}$.  A careful analysis shows that it cancels
between the two terms, rendering all gauginos massless.  (For more
discussion of RHS models, see \cite{dine macintire}.)

\subsection{Nonrenormalizable Hidden Sector Models}
\vspace{-0.25cm}

The prototypical NRHS model is motivated by superstring theory,
in which supersymmetry breaking is communicated by a dilaton field $S$.
(We ignore the additional complications from string moduli.  See, for
example, \cite{kaplunovsky louis}, and references therein.)  The
effective lagrangian contains terms of the form
$$ \int d^4\theta\ E\,\{ - M^2_P\,\log\left({ S + S^+\over M_P}
\right)\ +\ \Phi^+\Phi \} $$
\begin{equation}
{1\over M_P}\,\int d^2\theta\ \E\,S\,W^\alpha W_\alpha\ .
\label{dilaton-spurion}
\end{equation}

Now, in a NRHS model, the gauginos are assumed to condense, $\langle
\lambda^\alpha \lambda_\alpha \rangle = \langle W^\alpha
W_\alpha \vert \rangle$ $\simeq \Lambda^3$.  From (\ref{dilaton-spurion})
we see that this induces an expectation value of order $\Lambda^3/
M_P$ for the $\theta\theta$-component of $S$.  This, in turn, implies
that supersymmetry is spontaneously broken.

The great advantage of this scenario is that it is motivated by a very
general feature of string theory, the presence of a dilaton.  It induces
tree-level, weak-scale masses for the gauginos and the squarks and
sleptons.  The disadvantage is that the models are not calculable.  Typically,
there is no stable vacuum.  At best there is a cosmological solution, in which
the vacuum rolls off to the Planck scale, where it is assumed to be stabilized
by unknown string effects.  (See, however, \cite{racetrack}.)

\subsection{Common Features}
\vspace{-0.25cm}

At first glance, these three pictures differ considerably.  However,
for generic models -- that is, models whose superpotentials contain all
couplings allowed by symmetry -- they share a common feature:  the existence
of an extra, unwanted, global, U(1) symmetry, called $R$ symmetry \cite{nelson
seiberg}.  When supersymmetry is broken, the $R$ symmetry is also broken.
This gives rise to a massless Goldstone boson, called the $R$ axion.  Such
a particle is unacceptable for a variety of reasons. (See \cite{sikivie},
and references therein.)  The $R$ axion must be eliminated, one way or another.

One way to eliminate the $R$ axion is to construct models that are not generic.
In supersymmetric theories, this can actually be natural \cite{seiberg}, and
much
progress has been made in analyzing the properties of such theories
\cite{strong susy}.  Another escape is to add more structure so that
the $R$ axion becomes a pseudo-Goldstone boson \cite{dine nelson}. This
route leads to VS models that are quite complex.

In the rest of this talk, I will first explain why $R$ symmetry is associated
with generic models of dynamical supersymmetry breaking \cite{nelson seiberg}.
I will then show -- for VS and RHS models -- how supergravity effects
eliminate the problems with the $R$ axion \cite{bagger poppitz randall}.  I
will
conclude by illustrating how this works in the context of a specific model, the
original $3-2$ model of Affleck, Dine and Seiberg \cite{ads}.
This model provides a nice test case because the effects of dynamical
supersymmetry
breaking can be calculated in a controlled weak-coupling expansion.

\section{The Ubiquity of $R$ Symmetry}
\vspace{-0.25cm}

To understand the role of $R$ symmetry in dynamical supersymmetry breaking,
let us suppose that we have a supersymmetric theory whose coupling grows strong
at the scale $\Lambda$.  Let us consider the effective theory, valid below the
scale $\Lambda$.  Let us also assume that the effective theory is
supersymmetric, with the scale of supersymmetry breaking $M_S \lsim \Lambda$.

In this situation, the effective theory can be described by a K\"ahler
potential $K$ and superpotential $W$.  The superpotential is an analytic
function of the fields $z_i$, for $i = 1,...,\ n$.  Supersymmetry is
spontaneously broken if
\begin{equation}
{\partial W \over \partial z_i}\ \ne\ 0
\end{equation}
for some $i = 1,...,\ n$.

In general, the system
\begin{equation}
{\partial W \over \partial z_i}\ =\ 0
\label{no-susy-breaking}
\end{equation}
has a solution because it contains $n$ complex equations in $n$ complex
unknowns.  This implies that, for generic superpotentials, it is typically
not possible to break supersymmetry.

The situation does not change if $W$ is invariant under a $d$-dimensional
internal symmetry group.  In this case the superpotential
depends on $n-d$ complex variables, so (\ref{no-susy-breaking}) reduces to
$n-d$ complex equations in $n-d$ unknowns.  Generically, supersymmetry is
not broken.

Now, however, let us assume that the effective theory has a
spontaneously-broken continuous $R$ symmetry.  Under an $R$ symmetry, the
superpotential is not invariant, but has $R$-charge $-2$,
\begin{equation}
W\ \rightarrow\ e^{-2i\alpha}\,W\ .
\end{equation}
Since the $R$ symmetry is spontaneously broken, we are free to label our fields
so that the field $z_n$ has $R$-charge $q_n \ne 0$, with $\langle z_n \rangle
\ne 0$.  We can then write \cite{nelson seiberg}
\begin{equation}
W\ =\ z_n^{-2/q_n}\,F(x_j)\ ,
\end{equation}
where $x_j = z_j/z_n$, for $j = 1,...,\ n-1$.
In terms of the variables $x_j$, the conditions (\ref{no-susy-breaking}) become
$${\partial F \over \partial x_j} \ = \ 0 $$
\begin{equation}
F \ = \ 0\ .
\end{equation}
This system contains $n$ complex equations in $n-1$ unknowns.  It does not
generally have a solution, so supersymmetry is spontaneously broken
\cite{nelson
seiberg}.

This argument shows that -- for generic models -- spontaneously-broken $R$
symmetry is sufficient for dynamical supersymmetry breaking.  As discussed
above, the $R$ symmetry leads to a massless Goldstone boson -- the $R$ axion.
In what follows, we will show that supergravity effects explicitly break the
$R$ symmetry and give mass to the $R$ axion.

\section{The Supergravity Solution}
\vspace{-0.25cm}

In this section we shall see that supergravity provides a natural solution to
the problem of the $R$ axion \cite{bagger poppitz randall}.  To begin, we
recall that in rigid supersymmetry, the scalar potential $V$ can be written in
terms of the superpotential $W$ and the K\"ahler metric $K_{ij*}$,
\begin{equation}
V\ =\ K^{ij*}\,\partial_i W \, \partial_{j*} W^*\ ,
\label{global-potential}
\end{equation}
where, for simplicity, we ignore possible $D$-terms.
Supersymmetry is spontaneously broken if the vacuum energy is positive,
$\langle
\partial_i W \rangle \ne 0$, for some value of $i= 1,...,\ n$.

When supergravity is included, the scalar potential changes in the following
way,
\begin{equation}
V\ =\ \exp \left({K \over M^2_P}\right) \, \left[ K^{ij*}\,D_i W \, D_{j*} W^*\
-
\ {3\over M^2_P}\ W^*W \right]\ .
\label{local-potential}
\end{equation}
In this expression, the covariant derivative of the superpotential
is given by
\begin{equation}
D_i W\ =\ \partial_i W \ +\ {K_i\over M^2_P}\, W\ .
\end{equation}
The condition for supersymmetry breaking is $\langle D_i W \rangle
\ne 0$.

Comparing (\ref{global-potential}) with (\ref{local-potential}), we
see that all of the supergravity corrections are suppressed by powers
of $1/M_P$.  The only term of consequence is a possible constant in
the superpotential,
\begin{equation}
W\ \rightarrow\ W\ +\ c\ .
\end{equation}
In rigid supersymmetry, the constant has no effect and can
safely be ignored.  In local supersymmetry, however, the story is
different.  This is because the constant contributes to the vacuum
energy.  In fact, its role is to cancel the vacuum energy and
ensure that the cosmological constant is zero.  (In theories where all
scales are less than $M_P$, this is the {\it only} way to cancel the vacuum
energy.)

Given $\langle D_i W \rangle \simeq M^2_S$, the constant $c$ must
be of order $M^2_S M_P$ to cancel the vacuum energy.  Because
it grows with $M_P$, the constant is very important.  It cancels
the cosmological constant and generates the gravitino mass.   It induces
soft supersymmetry-breaking masses for the squarks and sleptons.  But
most importantly for this talk, it explicitly breaks the
$R$ symmetry and gives rise to an explicit mass
for the $R$ axion.

The mass of the $R$ axion is easy to find using a nonlinear realization.
Under a field-dependent $R$ transformation, the superpotential $W$ transforms
as follows,
\begin{equation}
W\ \ +\ c\ \rightarrow\ e^{2i A /f_A}\,W\ +\ c\ ,
\label{axiontransf}
\end{equation}
where $A$ is the axion field, and $f_A$ is the axion decay constant.
The mass of the axion can be found by substituting (\ref{axiontransf})
into (\ref{local-potential}) and expanding to second order in $A$.
One finds
\begin{equation}
M^2_A \ \simeq\ {c\over f^2_A\, M^2_P}\ \langle W \rangle\ ,
\end{equation}
which implies
\begin{equation}
M^2_A \ \simeq\ {M^2_S\, \Lambda^3 \over f^2_A\, M_P}\ .
\end{equation}

For VS models, $f_A \simeq \Lambda \simeq M_S$, so $M^2_A \simeq
\Lambda^3/M_P$.  Astrophysical constraints based on stellar and
supernova evolution require $M_A \gsim 10$ MeV \cite{raffelt},
which implies $\Lambda \gsim 10^5$ GeV.   Therefore VS models
are safe provided $\Lambda$ is above about $10^5$ GeV, as is
typically the case.

For RHS models, $f_A \simeq \Lambda \simeq M_S$, so $M^2_A \simeq
\Lambda^3/M_P$, as before.  Now, however, $M^2_S \simeq M_W M_P$, so
$M^2_A \simeq M_W M_S$, or $M_A \simeq 10^6$ GeV.  Such an axion is
too heavy to affect stellar dynamics.  It decays relatively quickly
so it is also cosmologically safe.  (See \cite{bagger poppitz randall},
and references therein.)  Therefore RHS models do not have problems
with the $R$ axion.

Finally, for NRHS models, $f_A \simeq M_P$ and $M^2_S \simeq \Lambda^3/M_P$.
For such models, $M_A \simeq M_W$, about 100 GeV.  Such a light, weakly coupled
axion is cosmologically dangerous, but no more so than the light moduli
fields that arise in string theory \cite{cosmo moduli}.  Presumably,
the mechanism that cures the cosmological moduli problem also cures the
cosmological problem with the $R$ axion.
(For recent progress in this direction, see \cite{cosmo moduli2}.)

\section{The $3-2$ Model}
\vspace{-0.25cm}

The so-called $3-2$ model of Affleck, Dine and Seiberg provides a classic
example
in which dynamical supersymmetry breaking is realized in a controlled,
weak-coupling
expansion \cite{ads}.  The model can serve as a prototypical RHS theory, or,
suitably generalized, as the basis of a VS theory of dynamical supersymmetry
breaking \cite{dine nelson shirman}.

\subsection{The Model}
\vspace{-0.25cm}

\renewcommand{\arraystretch}{1.2}
\begin{table}[t]
\begin{center}
\caption{The fields of the $3-2$ model.}
\vspace{.25 truein}
\begin{tabular} {|c|c|c|c|c|} \hline
Particle &\ \ SU(3) $\times$ SU(2) \ \
&\ \ Hypercharge\ \ &
\ \ $R$-charge\ \ \\\hline
$Q$ & (3, 2) & 1/6 & 1 \\
$\bar U$ & ($\bar 3$, 1) & $-2/3$ & 0 \\
$\bar D$ & ($\bar 3$, 1) & 1/3 & 0 \\
$L$ & $(1, 2)$ & $-1/2$ & $-3$ \\\hline
\end{tabular}
\end{center}
\end{table}
\renewcommand{\arraystretch}{1.0}

The $3-2$ model is based on two-flavor supersymmetric QCD, with
a gauged SU$(2)_L$ flavor symmetry.  To describe the
model, let us denote the left- and right-handed quark superfields
as $Q$ and $\bar{Q} = ( \bar U,\bar D)$.  Under the SU(3) $\times$
SU(2) gauge symmetry, the quark superfields transform as shown
in Table 1.

Note that the particle content of the model is similar to that of
the minimal supersymmetric standard model, without the Higgs and the
right-handed electron superfields.  Apart from the gauge symmetries,
the model also has two anomaly-free continuous global symmetries:
hypercharge, U$(1)_Y$, and an $R$ symmetry, U$(1)_R$.  The hypercharge
and $R$-charge assignments are also listed in Table 1.

\renewcommand{\arraystretch}{1.2}
\begin{table}[t]
\begin{center}
\caption{The fields of the effective theory.}
\vspace{.25 truein}
\begin{tabular} {|c|c|c|c|c|} \hline
Particle &\ \ Hypercharge\ \ &
\ \ $R$-charge\ \ \\\hline
$X_1$ & 0 & $-2$ \\
$X_2$ & $-1$ & $-2$ \\
$X_3$ & $0$ & $2$ \\\hline
\end{tabular}
\end{center}
\end{table}
\renewcommand{\arraystretch}{1.0}

The K\"ahler potential of the model takes the usual form
\begin{equation}
K \ =\  Q^+ Q
\ +\ \bar{Q} \bar{Q}^+\ +\ L^+ L \ .
\label{kahler}
\end{equation}
In (\ref{kahler}) the SU(2) and SU(3) gauge superfields are not
written, but are assumed to be coupled in the usual way \cite{wess
bagger}.

In the absence of a superpotential, the scalar potential vanishes
for a number of flat directions in field space.  Therefore the ground
state is undetermined at the classical level \cite{ads}.  The equations
that determine the flat directions are
\begin{equation}
Q^{+m} Q_{\ell}\ - \ \bar{Q}^{m} \bar{Q}^{+}_{\ell}\ =\  0
\label{flatdirectionsSU3}
\end{equation}
for the SU(3) $D$-terms, and
\begin{equation}
Q^+_{\alpha}Q^{\beta} \ +
\ L^{+}_{\alpha}L^{\beta}\ =\ {1\over2}~\delta^{~\beta}_{\alpha}
{}~( Q^{+} Q ~+~ L^{+} L )
\label{flatdirectionsSU2}
\end{equation}
for the SU(2) $D$-terms.  Up to local symmetries, the solutions to
these equations are parametrized by six real variables, also called
moduli.  The moduli parametrize inequivalent, supersymmetry-preserving
vacua.  The variations along the moduli directions correspond to six real,
massless scalar fields.

Let us now consider the theory expanded around a solution of
(\ref{flatdirectionsSU3}), (\ref{flatdirectionsSU2}), such that the
vacuum expectation values $v$ obey
\begin{equation}
v \ \gg\ \Lambda \  ,
\label{scale}
\end{equation}
where
\begin{equation}
\Lambda \ =\ v\  \exp \left(-{ 8\pi^2\over g(v)^2~b_0}\right)\ .
\label{Lambda3}
\end{equation}
In this expression,
$\Lambda$ is the scale where the SU(3) gauge coupling $g$ becomes strong,
and $b_0$ is the one-loop coefficient of the SU(3) beta function.  For $v
\gg \Lambda$, the theory is in the weak-coupling regime.  The SU(3) and SU(2)
gauge symmetries are completely broken so the vector supermultiplets are
massive.  Supersymmetry is unbroken, so 11 out of the 14 matter chiral
superfields are absorbed by the vector superfields.  The six real moduli
are contained in three massless chiral superfields.

At energies below the scale $\Lambda$, the low-energy effective theory
can be described in terms of three gauge-invariant
chiral superfields,
\begin{eqnarray}
\label{lightfields}
X_1 &=& Q\, \bar{D}\, L\;,\nonumber \\
X_2 &=& Q\, \bar{U}\, L\;,\;\\
X_3 &=& \det~ \bar{Q}_{\alpha}\;Q^{\beta} \; ,\nonumber
\end{eqnarray}
whose scalar components parametrize the six massless moduli.  The
quantum numbers of these fields are listed in Table 2.

\begin{figure}[t]
\epsfysize=2.5in
\hspace*{0.9in}
\epsffile{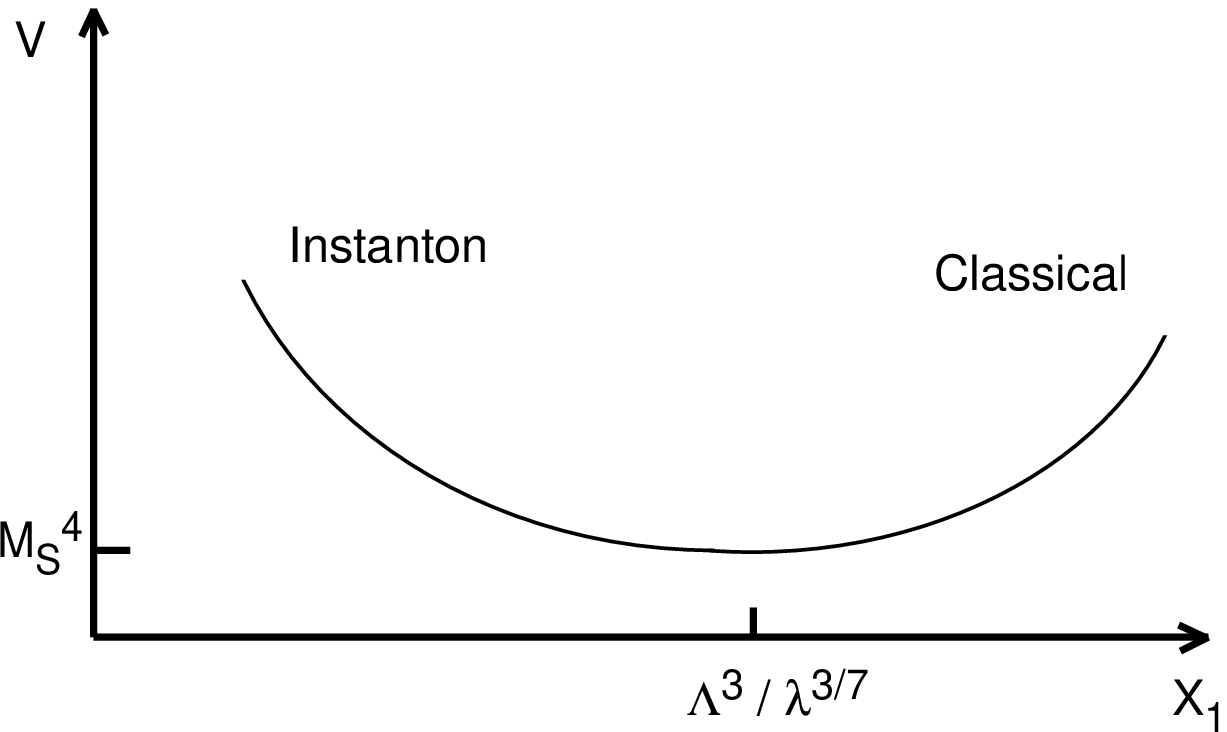}
\fcaption{The effective potential in the $3-2$ model.}
\end{figure}

Let us now discuss the superpotential of the effective theory.  The
vacuum preserves the global hypercharge symmetry, so we take the superpotential
to preserve it as well,
\begin{equation}
W \ =\  \lambda \; X_1\ +\  2 \; {\Lambda^7 \over X_3}\ .
\label{superpotential}
\end{equation}
The first term is the renormalizable superpotential that we assume to be
present in the classical theory.  The second is nonrenormalizable, and can
be shown to be generated by nonperturbative effects.  Its coefficient can be
calculated in a weak coupling expansion around a constrained instanton
vacuum \cite{ads}.  Note that this superpotential also respects U(1)$_R$.
The $R$ symmetry is ``accidental" in the sense that it is a direct result
of hypercharge conservation in the effective theory.

In the presence of the superpotential, the scalar potential is no
longer flat.  Indeed, when $\lambda = 0$, the scalar potential does not
have a minimum and the theory does not have a well-defined ground
state \cite{ads}.  For the case when
\begin{equation}
\lambda \ \ll\ g_2 \ \ll\ g_3 \ \ll\ 1 \ ,
\label{lambda}
\end{equation}
the potential has a minimum at finite values for the fields, of order
\begin{equation}
v \ \simeq \ {\Lambda\over\lambda^{1/7}}\ ,
\label{vevs}
\end{equation}
as shown in Figure 1.  This value is such that the weak coupling assumption
(\ref{scale}) is self-consistent, so the theory can be analyzed perturbatively.

As we will see, the vacuum energy is positive and supersymmetry is
spontaneously broken, with $M_S \simeq \lambda^{5/14} \Lambda$.  The
moduli are massive, with masses of order $\lambda^{6/7} \Lambda$.  The
hierarchy of scales is summarized in Figure 2.

\begin{figure}[t]
\epsfysize=3.5in
\hspace*{1.3in}
\epsffile{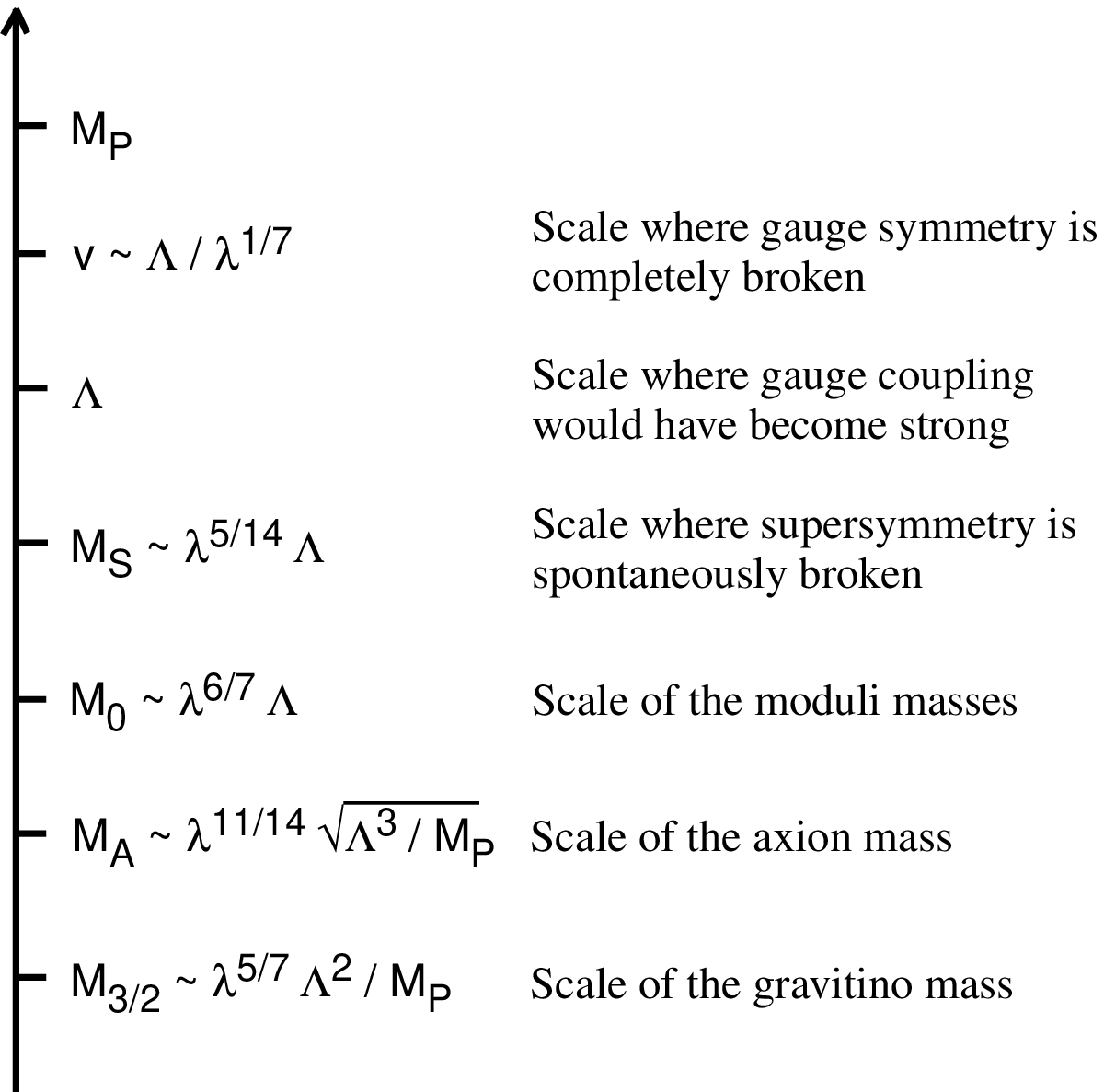}
\fcaption{The mass scales of the $3-2$ model.}
\end{figure}

\subsection{The Low-Energy Sigma Model and its Spectrum}
\vspace{-0.25cm}

In this section we will study the effective field theory below the
scale $\Lambda$.  We will find the spectrum of all particles lighter than
this scale.

In the limit (\ref{lambda}), the K\"ahler potential of the effective theory
is given by the projection of (\ref{kahler}) onto the moduli fields $X_1, X_2$
and $X_3$. (See also \cite{poppitz randall}.)  Using the notation of ref.
\cite{ads}, the resulting K\"ahler potential can be written
\begin{equation}
K\  = \ 24\ { A \ + \ B x \over x^2} \ ,
\label{sigmak}
\end{equation}
where
$$ A \ =\ {1\over 2}\ ( X^{+}_1 \, X_1 \ =\ X^{+}_2\, X_2 ) $$
\begin{equation}
B \ = \ {1\over 3}\ \sqrt{X^{+}_3 ~X_3} \ ,
\label{AB}
\end{equation}
and
\begin{equation}
x \ =\ 4~\sqrt{B}\; {\rm cos} \left(~{1\over 3}
\; {\rm Arccos} ~ {~A~ \over B^{3/2}}~\right) \; .
\label{eks}
\end{equation}
The equations that determine $x$ as a function of the light
superfields have several solutions.  Equation (\ref{eks}) is the only one
that leads to a positive definite K\"ahler metric at the minimum.

The low-energy theory is therefore described by a sigma model with K\"ahler
potential $K$ and superpotential (\ref{superpotential})
\begin{equation}
\label{sigmaw}
W\ = \ \lambda\; X_1 \ + \ 2\ {\Lambda^7 \over X_3}\; .
\end{equation}
To find the ground state, one must minimize the scalar potential.
After a numerical analysis, one finds the following values for the expectation
values of the moduli fields:
\begin{eqnarray}
X_1  &=&  0.50\; \Lambda^3 / \lambda^{3/7} \nonumber \\
X_2  &=&  0 \nonumber \\
X_3  &=&  2.58\; \Lambda^4 / \lambda^{4/7}\ .
\label{sigmamin}
\end{eqnarray}
At the minimum, the vacuum energy density is
\begin{equation}
M_S^4 \ =\ 3.59\;\lambda^{10/7}\; \Lambda^4 \ .
\label{vacenergy}
\end{equation}

The scalar mass matrix is found by expanding the potential about its
minimum.  It is given by
\begin{equation}
M^2_{ab} \ =\  \langle V_{ab} \rangle\ ,
\label{VVVV}
\end{equation}
where $V$ is the scalar potential, and $a,b = 1,...,6$
label the six light real fields.  One discovers three real scalar fields of
masses $3.88,\ 2.83\ {\rm and} \ 2.04$ (in units of $\lambda^{6/7}\Lambda$),
one complex scalar of mass $1.35$ (in the same units), and a massless $R$
axion \cite{bagger poppitz randall}.

The fermion mass matrix is \cite{wess bagger}
\begin{equation}
M_{ij}\ =\ \langle W_{ij}\ -\ K_{k\ell *}^{-1}~K_{ij\ell *}\, W_k \rangle\ ,
\label{fermionmasssigma}
\end{equation}
where $i,j = 1,...,3$ label the three light fermions.  One finds
a massless goldstino, a massless fermion of hypercharge one, and a
fermion of mass $3.19 \; \lambda^{6/7}\; \Lambda$  \cite{bagger poppitz
randall}.

\subsection{Supergravity Couplings and $R$ Axion Mass}
\vspace{-0.25cm}

In this section we couple the $3-2$ model to supergravity and compute the
supergravity contribution to the $R$ axion mass.  The supergravity coupling
is straightforward and can be done using the effective theory of the previous
section.

As discussed above, in any supergravity theory where all scales are smaller
than $M_P$, the vacuum energy is canceled by adding a constant
to the superpotential,
\begin{equation}
c\ =\ {1\over \sqrt{3}}\ M_S^2 M_P \ .
\label{constantW}
\end{equation}
In this class of models, all soft breaking terms
are induced by $c$.  For the $3-2$ model, gravitino
mass is
\begin{equation}
M_{3/2}\ =\ { c \over M_P^2}\ +\
{1\over \sqrt{3}}~{M_S^2 \over M_P}\ =\
1.09~ {\lambda^{5/7} \; \Lambda^2\over M_P} \ .
\label{m3/2}
\end{equation}

For our purposes, the most important consequence of the constant $c$ is
the fact that it explicitly breaks the $R$ symmetry.  In particular, it
induces $R$-symmetry-breaking terms in the supergravity-coupled scalar
potential, including
\begin{equation}
V' \ =\ {1\over \sqrt{3}}~{M_S^2 \over M_P}\left(W_{i}\,K_{ij*}^{-1}\, K_{j*}~
-~3\,W \right)
\ +\ {\rm h.c.} \ +\  ...\ ,
\label{v1}
\end{equation}
where $K$ and $W$ are the K\"ahler potential (\ref{sigmak}) and superpotential
(\ref{sigmaw}) of the effective theory, and the dots denote terms suppressed by
additional powers of $M_P$.

As discussed above, these terms give mass to the $R$ axion.
The axion coupling constant turns out to be
\begin{equation}
f_A\ =\  2.18~{\Lambda \over \lambda^{1/7}}\ =\ 1.58 ~{M_S \over
\sqrt{\lambda}}\ .
\label{faxion}
\end{equation}
The axion mass is then \cite{bagger poppitz randall}
\begin{equation}
M_A^2\ =\ 10.0 \; \lambda^{11/7}\; {\Lambda^3 \over M_P} \ =\
6.58 \sqrt{\lambda}~ M_{3/2}\; M_S \ .
\label{rmass}
\end{equation}
This formula in is agreement with our previous results for VS and RHS models.

\section{Conclusions}
\vspace{-0.25cm}

In this talk we have seen that models of dynamical supersymmetry breaking
offer an attractive explanation for the ratio $M_W/M_P \simeq 10^{-17}$.
We have seen why it is difficult to construct such models, and demonstrated
why many candidate models contain a potential $R$ axion.

In VS models with a supersymmetry breaking scale greater
than about $10^5$ GeV, the axion is sufficiently heavy to evade astrophysical
constraints.  In RHS models, the axion mass is quite large, of order
$10^6$ GeV, so the axion is astrophysically and cosmologically safe.  In NRHS
models, the axion mass is of order 100 GeV.  Such a light, weakly-coupled axion
can lead to cosmological difficulties of the sort already present for the
moduli fields of string theory.

I would like to thank Roberto Casalbuoni and Luca Lusanna for organizing
this delightful conference in the Villa Spelman, and my collaborators
Erich Poppitz and Lisa Randall for their many insights into the work
presented here.   This work was supported by the U.S. National Science
Foundation under grant NSF-PHY-9404057.

\section{References}
\vspace{-0.25cm}

\nc{\ib}[3]{ {\it ibid.\ }{\it #1} (19#2) #3}
\nc{\np}[3]{ {\it Nucl.\ Phys.\ }{\it #1} (19#2) #3}
\nc{\pl}[3]{ {\it Phys.\ Lett.\ }{\it #1} (19#2) #3}
\nc{\pr}[3]{ {\it Phys.\ Rev.\ }{\it #1} (19#2) #3}
\nc{\prep}[3]{ {\it Phys.\ Rep.\ }{\it #1} (19#2) #3}
\nc{\prl}[3]{ {\it Phys.\ Rev.\ Lett.\ }{\it #1} (19#2) #3}

\end{document}